\documentclass[10pt, twocolumn, fleqn]{article}

\usepackage[utf8]{inputenc}
\usepackage{amsmath, amssymb, mathtools, physics, bm}
\usepackage[low-sup]{subdepth}

\usepackage[letterpaper, left=0.4in, right=0.4in, top=1in, bottom=1in]{geometry}
\usepackage[labelfont=bf]{caption}
\usepackage{hyperref}  % Liens
\hypersetup{colorlinks, urlcolor=blue, citecolor=blue}
\usepackage{titlesec}
\titleformat{\section}{\large\bfseries}{\thesection.}{10pt}{\large}
\titleformat{\subsection}[runin]{\normalsize\bfseries}{\thesubsection}{8pt}{\normalsize}
\newcommand{\para}[1]{\subsection{#1.}}

\newcommand{\eq}[1]{\begin{equation} #1 \end{equation} }
\newcommand{\eqa}[1]{\begin{align}\begin{split} #1 \\[5pt] \end{split} \end{align}}  
\newcommand{\eqn}[1]{\begin{equation*} #1 \end{equation*} }

\newcommand{\Hu}{\mathcal{H}}
\newcommand{\p}{\partial}
\newcommand{\cov}{\nabla}
\newcommand{\A}{\mu}  % pour les indices des tenseurs
\newcommand{\B}{\nu}
\newcommand{\C}{\rho}

\renewcommand{\expval}[1]{\langle #1 \rangle}

\DeclareMathOperator{\Ci}{Ci}

\usepackage{abstract}

\usepackage{color,soul}
\usepackage{cite}
\usepackage{ftnright}

\begin{document}

\twocolumn[{ 
\centering
\textbf{\Large Perturbative Correction to the Average Expansion Rate \\ of Spacetimes with Perfect Fluids}

\vspace{15pt}
\large Vincent Comeau\footnotemark \vspace{5pt}

\textit{Department of Physics -- McGill University} \\
(\today)
\vspace{15pt}

\begin{abstract}
This paper discusses the leading-order correction induced by cosmological perturbations on the average expansion rate of an expanding spacetime, containing one or many perfect fluids. The calculation is carried out up to the second order in the perturbations, and is kept as general as possible. In particular, no approximation such as a long-wavelength or a short-wavelength limit is invoked, and all three types of perturbations (scalar, vector, and tensor) are considered. First, the average value of the expansion rate is computed over a three-dimensional space-like surface where the total density of the fluids is constant. Then, a formula is derived relating that average value to the one over any other surface, on which a different scalar property of the fluids is constant. Moreover, the general formulas giving the correction to the average expansion rate are applied, in particular, to the case of a spacetime containing a single fluid with a constant equation of state. The sign and the effective equation of state of the corresponding back-reaction effect in the first Friedmann equation are examined.
\vspace{15pt}
\end{abstract}
}]

\footnotetext{Email: \href{mailto:vincent.comeau@mail.mcgill.ca}{vincent.comeau@mail.mcgill.ca}}

{ \hypersetup{hidelinks} \tableofcontents }

\newpage
\section{Introduction}

\para{Goal of the paper}

According to the principles of modern cosmology, our universe can be modelled as an expanding spacetime, whose expansion rate changes with time. Such an expansion rate represents an important parameter in its own right, but even more so today, since its current value has yet to be determined accurately. There still exists a discrepancy between the local measurements of the expansion rate, for instance using standard candles, and the measurements relying on phenomena from the early universe, such as the cosmic microwave background. The latter systematically yield a value for the expansion rate smaller than the former. A possible explanation for this ``Hubble tension'' could come from the effects of spacetime inhomogeneities on the value of the expansion rate.

Inhomogeneities could also play a role in explaining the late-time acceleration of the expansion of our universe. In standard cosmology, this acceleration is attributed to a cosmological constant, or more generally to an additional component presumably contained in the universe called ``dark energy'', whose nature is still unknown. As contemplated for instance in \cite{Rasanen:2003, Barausse:2005, Geshnizjani:2005, Hirata:2005}, this dark energy might be in fact the result of nonlinear effects induced by the matter inhomogeneities. 

Similarly, it has been suggested that the long-wavelength modes of inhomogeneities in the early universe could lead to a dynamical relaxation of a bare positive cosmological constant, leaving behind a remnant which could explain the present-time acceleration. Both cosmological perturbations and gravitational waves could contribute to this effect \cite{Brandenberger:2002, Tsamis:1996a, Tsamis:1996b}.

The purpose of this paper is not to ascertain whether the back-reaction of inhomogeneities on the expansion rate really does represent a possible solution of the Hubble tension and dark energy problems. Our more modest goal is to compute the leading-order effect of cosmological perturbations on the spatial average of the expansion rate, for a spacetime containing various perfect fluids possibly moving at different velocities, and having different equations of state.

Several papers have presented similar calculations, but under more restrictive assumptions. For instance, in \cite{Kolb:2004}, the average expansion rate is computed for a spacetime containing pressureless ``cold'' matter, whereas in \cite{Geshnizjani:2002, Geshnizjani:2003} they consider the case of a spacetime with one or two scalar fields, in the long-wavelength limit.

Here, we consider more generally a spacetime containing perfect fluids, without making any assumptions regarding their respective equations of state or the way they interact with one another. We compute the average expansion rate of such a spacetime up to the second order in the metric perturbations. In doing so, we do not make use of any kind of approximation, such as a long-wavelength or a short-wavelength limit. All second-order terms are kept in our calculation, including all the time and spatial derivatives of the perturbations which contribute at that order. However, we do assume that the perturbations obey periodic conditions at the boundary of the volume over which the expansion rate is averaged.

Hence, by keeping our calculation as general as possible, we obtain formulas for the average expansion rate which can be applied to all epochs of our universe's history, at both early and late times, and for all types of perturbations. On the other hand, most previous papers on this topic only focused on specific epochs, such as the early inflationary stage, or the late-time matter-dominated period.

\para{Gauge issue and cosmic clock}

In this paper, we consider an expanding homogeneous background, to which are added small perturbations. However, such a distinction between the background and its perturbations bears no physical significance in the context of general relativity. The background is a pure mathematical construction which differs from the actual spacetime, and the perturbations added unto it can be altered arbitrarily by infinitesimal coordinate transformations. Hence, the results of any perturbative calculation of this kind, especially when it is carried out beyond the linear order, must be interpreted with care.

The issue at play here lies with the choice of coordinates. According to the principle of relativity, coordinates are mere labels for the points in spacetime, and as such they can be modified arbitrarily without affecting the physics of a phenomenon. Coordinates alone do not bear any physical reality, unless they can be related to a measurable property of the matter fields contained in the spacetime. Then, and only then, do they represent more than mere labels.

For instance, in cosmology, when the universe is modelled as homogeneous and expanding, most of the time coordinates which could be used to label points in spacetime are devoid of any physical meaning. Local observers would not be able to synchronize the clocks associated with these time coordinates because of their motion relative to one another, due to the spacetime expansion. Only the Friedmann-Lema\^itre time coordinate bears a clear physical meaning, as it can be related to the total fluid density. Only for the Friedmann-Lema\^itre time coordinate can local observers synchronize their clocks, by measuring the fluid density at the point where they are located.

Similarly, in the presence of cosmological perturbations, the time coordinate of the metric should be related to a measurable property of the matter fields, otherwise it becomes impossible to interpret the results of perturbative calculations involving a spatial average. For instance, the time coordinate could be expressed as a function of the density of a fluid contained in the spacetime, or in terms of the temperature of a field, such as the cosmic microwave background. Local observers simply need to agree initially on their common ``instrument'', on what specific property of matter they will be using to track time.

\para{Previous work}

The effects of inhomogeneities on the spatial average of the expansion rate have been previously considered following two main approaches, either perturbatively as in this paper, or non-perturbatively. The latter approach, which was pioneered by \cite{Buchert:1999, Buchert:2001}, involves taking the spatial average of the scalar components of the Einstein equations, and then comparing the resulting equations to their counterparts in a purely homogeneous spacetime. The main drawback of this method is that the set of ordinary differential equations to which it leads cannot be closed, and thus requires additional assumptions to be solved.

The effects of inhomogeneities have also been studied perturbatively, generally up to the second order in the metric perturbations. For instance, in \cite{Abramo:1997}, the average energy-momentum tensor induced at the second order by the perturbations is calculated in the Poisson gauge, for a spacetime containing a single scalar field. In the long-wavelength limit and for scalar perturbations, this second-order energy-momentum tensor emulates that of a negative cosmological constant. However, as pointed out in \cite{Unruh:1998}, this effect can be made to vanish by an appropriate second-order coordinate transformation.

The main issue with the result we just described is not that it depends on a specific gauge choice, or that it could be made to vanish if another gauge were used. Indeed, these arguments apply equally well to any perturbative calculation, including the one presented in \cite{Unruh:1998} as a disproof of the aforementioned result! The main reason why the latter is incorrect is that the Poisson gauge does not bear a clear physical interpretation in the long-wavelength limit. Although mathematically correct, its significance is physically unclear, since the Poisson gauge can be properly interpreted only in the short-wavelength limit, when the laws of general relativity resemble that of Newton. 

A better choice would be to use, for instance, a gauge in which the scalar field is equal to its background value, as argued in \cite{Abramo:1998, Afshordi:2000, Geshnizjani:2002}. Such a gauge provides a clear physical meaning for the surfaces over which the energy-momentum tensor is averaged, as these surfaces correspond to constant values of the scalar field. In that case, the back-reaction of scalar perturbations vanishes in the long-wavelength limit, rather than behaving as a cosmological constant.

Similarly, in \cite{Kolb:2004}, the leading-order correction to the average expansion rate is computed in the synchronous comoving gauge, for a spacetime containing cold matter. The expansion rate is averaged over the three-dimensional surfaces normal to the matter flow. Such a slicing is well-defined as long as the vector components of the perturbations can be neglected. As argued in \cite{Adamek:2017}, there comes a point in the evolution of the perturbations where the matter flow acquires significant vorticity, and thus the synchronous comoving gauge breaks down\footnote{The authors of \cite{Adamek:2017} also argue that the real question is not whether there exists slicings for which the back-reaction of perturbations is large, but whether there are some where it is small. Their remark is partially correct: it should be added that the chosen slicing must also bear a clear physical interpretation.}. Beyond the linear order, the scalar and tensor perturbations start to act as sources for the vector perturbations, which can then no longer be neglected. Therefore, although the synchronous comoving gauge bears a clear physical meaning, it can be used only under certain circumstances, which is why we chose not to use it in this paper.

\para{Summary of the results}

In section \ref{section-framework}, we state some preliminary results which will be used in the rest of the paper, and we specify our notation. We then compute the average value of the expansion rate over surfaces where the total density of the fluids is constant, first in full generality in section \ref{section-ave-total}, and for specific kinds of fluids in section \ref{section-applications}. Finally, in section \ref{section-ave-partial}, we consider the case when the average is taken instead over surfaces where some other scalar property of the fluids is constant, rather than its total density. The average value of the expansion rate on the former slicing is related to its value on the latter.

The general results of this paper are given in equations (\ref{scalar-correction}), (\ref{vector-correction}), and (\ref{tensor-correction}), which represent the leading-order correction to the average expansion rate over surfaces of constant total density. Another important result is (\ref{diff-two-averages}), which connects the latter correction to the one over an alternative surface, for a different clock field. 

Moreover, the table on the right-hand side of this page presents the behaviour of the corrections induced by the scalar, vector, and tensor perturbations respectively, for spacetimes containing a single fluid with a constant equation of state $w$. These corrections affect the first Friedmann equation written in terms of the average expansion rate, by supplying an additional density whose equation of state $w_\text{br}$ at early and late times is also provided in the table. In these cases, the early-time and late-time limits coincide respectively with the long-wavelength and short-wavelength limits of the perturbations.

The results derived in section \ref{section-ave-partial} of this paper have been recently discussed in \cite{Comeau:2023a}, and applied to the particular case of a spacetime containing at least two fluids, one of which is used as a clock. The expansion rate is averaged over surfaces where the density of the clock fluid is constant. It is shown that the long-wavelength modes of the perturbations lead to a decrease of the average expansion rate compared to its background value, which confirms the analysis of \cite{Geshnizjani:2002, Marozzi:2012, Brandenberger:2018}.\vspace{0pt}

\begin{figure}
\begin{center}
\begin{tabular}{|ccc|}
\hline
  & Early time & Late time \\ 
  & (Super-Hubble) & (Sub-Hubble) \\[2pt]
\hline
 Scalar perturbations & Negative & Negative \\
 $w = 0$ & $w_\text{br} = -\frac13$ & $w_\text{br} = -\frac23$ \\[2pt]
\hline
 Scalar perturbations & Negative & Positive \\ 
 $w > 0$ & $w_\text{br} = -\frac13$ & $w_\text{br} = -w$ \\[2pt] 
\hline
 Tensor perturbations & Positive & Positive \\ 
 $\frac13 < w < 1$ & $w_\text{br} = -\frac13$ & $w_\text{br} = \frac13$ \\[2pt] 
\hline
 Tensor perturbations & Positive & Positive \\ 
 $-\frac13 < w < \frac13$ & $w_\text{br} = -\frac13$ & $w_\text{br} = w$ \\[2pt] 
\hline
 Vector perturbations & Positive & Positive \\ 
 $ w > \frac13$ & $w_\text{br} = 1$ & $w_\text{br} = -w + \frac23$ \\[2pt] 
\hline
 Vector perturbations & \multicolumn{2}{c|}{Decays with time} \\ 
 $ w < \frac13$ & \multicolumn{2}{c|}{} \\[2pt] 
\hline
\end{tabular}
\end{center}
\textit{Figure:} Corrections to the average expansion rate induced by the scalar, vector, and tensor perturbations respectively, for a spacetime containing a single fluid with a constant equation of state $w$. Here, $w_\text{br}$ denotes the effective equation of state describing the back-reaction of perturbations.\vspace{-10pt}

\end{figure}

\section{Framework} \label{section-framework}

\para{Energy-momentum of fluids}

In this paper, we consider a slightly inhomogeneous spacetime, with a metric $g_{\A\B}$, containing one or many perfect fluids. Each of the fluids is labelled with a different index $A$, has a density $\rho_A$, a pressure $p_A$, and moves through the spacetime with a normalized velocity $u_A^\A$,
\eq{
g_{\A\B} u_A^\A u_A^\B = -1\,.
}

The total density and pressure of the fluids in the spacetime are thus
\eqa{
\rho &= \sum_A \rho_A\,, \\[1pt]
p &= \sum_A p_A\,,
}
where the sums are over all the fluids contained in the spacetime. The total energy-momentum tensor is given by the sum of the energy-momenta from each fluid,
\eq{ \label{energy-momentum}
T_{\A\B} = \sum_A \Big((\rho_A + p_A) \,u_\A^A \,u_\B^A + p_A\,g_{\A\B}\Big)\,.
}

Since we want our calculation to remain as general as possible, we will not make any further assumptions regarding the fluids contained in the spacetime. Specific cases will be considered later, in section \ref{section-applications}, after the main result has been obtained.

\para{Expansion rate}

The main goal of this paper is to compute the average expansion rate of the spacetime over surfaces of constant density. In general, an expansion rate can be computed by tracking an ensemble of particles present everywhere in the spacetime, and moving along trajectories $x^\A = x^\A(T)$, where $T$ is the proper time of the particles. Their velocity is then
\eq{
U^\A = \dv{x^\A}{T}\,.
}

Since $T$ is chosen to be the proper time of the particles, their velocity is normalized to $g_{\A\B} U^\A U^\B = -1$. The expansion rate $H$ at a given point can be defined by tracking a small volume of particles in the vicinity of that point. Assuming the small volume of particles $\delta V$ has an average size $\delta L$, with $\delta V = (\delta L)^3$, the expansion rate is then
\eq{
H = \frac{1}{\delta L}\dv{(\delta L)}{T} = \frac{1}{3\,\delta V}  \dv{(\delta V)}{T}\,.
}

As shown in \cite{Ehlers:1993}, it can also be written in terms of the velocity of the particles,
\eq{ \label{def-expansion-rate}
H = \frac13 \cov_\A U^\A\,.
}
where $\cov_\A$ denotes the covariant derivative. This expression reduces to the usual Hubble expansion rate when the spacetime is homogeneous. 

In the first part of this paper, we will not make any assumption regarding the set of particles (and their velocity $U^\A$) chosen to measure the expansion rate. We will refer to them, generally speaking, as the ``observed particles''. They could be, for instance, free observers falling along geodesics, or one of the fluids contained in the spacetime, moving at a velocity $u_A^\A$. Each of these can be used to define a different expansion rate for the spacetime, which might be more or less appropriate depending on the measurement we intend to make. For example, when measuring the expansion rate at late times, from the receding motion of galaxies at low redshift, the velocity $U^\A$ should be identified with the cold matter component of our universe, which includes the galaxies whose motion we measure.

\para{Metric perturbations}

The spacetime considered in this paper can be described as a homogeneous background, onto which are added small perturbations. The background is spatially flat, expanding with time, and has a metric which can be written as $g_{\A\B} = a^2(\eta)\,\eta_{\A\B}$. Here, $a(\eta)$ is the scale factor, $\eta$ is the conformal time, and $\eta_{\A\B}$ is the Minkowski metric with the signature $\eta_{00} = -1$. The background expansion rate is thus
\eq{ \label{background-expansion-rate}
H_0 = \frac{\Hu}{a} = \frac{\dot{a}}{a^2}\,,
}
where a dot denotes a derivative with respect to the conformal time. Throughout this paper, the subscript 0 indicates either the time component of a vector, or the background value of a scalar quantity (and not its value at present time). When small perturbations are added, the metric becomes
\eq{
g_{\A\B} = a^2\,\hat{g}_{\A\B} = a^2\big( \eta_{\A\B} + h_{\A\B}\big)\,,
}
where $h_{\A\B}$ represents the metric perturbations. Up to the second order in the perturbations, the inverse components of the metric are
\eq{
g^{\A\B} = \frac{1}{a^2}\,\hat{g}^{\A\B} = \frac{1}{a^2}\big( \eta^{\A\B} - h^{\A\B} + h^\A_\C h^{\B\C} \big)\,.
} 

The indices of the perturbations are raised and lowered using the Minkowski metric, $h_\A^\B = \eta^{\B\C} h_{\A\C}$. In these formulas, $\hat{g}_{\A\B}$ represents an auxiliary metric, related to the actual spacetime metric through a scale transformation. Indeed, it is often worth applying a scale transformation of the metric to get rid of the background scale factor, before expanding quantities in series. Similarly, we can define an auxiliary velocity,
\eq{
\hat{U}^\A = a\,U^\A\,,
}
which is normalized with respect to the auxiliary metric, $\hat{g}_{\A\B} \hat{U}^\A\hat{U}^\B = -1$, and is thus constant in the background, $\hat{U}^0 = 1$ and $\hat{U}^i = 0$. After such a scale transformation, the expansion rate (\ref{def-expansion-rate}) becomes
\eqa{
\frac{H}{H_0} &= \hat{U}^0 + \frac{1}{3\Hu} \hat{\cov}_\A \hat{U}^\A \\[4pt]
&= \hat{U}^0 + \frac{1}{3\Hu}\left( \p_\A \hat{U}^\A + \frac12 \hat{g}^{\A\B} \p_\C \hat{g}_{\A\B}\, \hat{U}^\C\right)\,.
}
Expanding up to the second order, we get
\eq{ \label{expansion-rate-series}
\frac{H}{H_0} = \hat{U}^0 + \frac{1}{3\Hu}\left( \p_\A \hat{U}^\A + \frac12 \p_\A h\,\hat{U}^\A - \frac12 h^{\A\B} \dot{h}_{\A\B} \right)\,,\,\,\,
}
where $h=h_\A^\A$ is the trace of the metric perturbations. The velocity of the observed particles can also be expanded in series. Up to the first order, the spatial velocity is
\eq{ 
\hat{U}^i = \hat{g}^{i\A} \,\hat{U}_\A = \hat{U}_i - h_{0i} \,.
}

Using the normalization condition $\hat{g}_{\A\B} \hat{U}^\A\hat{U}^\B = -1$, we get at the second order,
\eqa{ \label{obs-velocity-time}
\hat{U}^0 &= 1 + \frac12\phi + \frac38 \phi^2 + h_{0i} \hat{U}^i + \frac12 \hat{U}^i \hat{U}^i 
\\[4pt] &= 1 + \frac12 \phi + \frac38 \phi^2 - \frac12 h_{0i} h_{0i} + \frac12 \hat{U}_i \hat{U}_i\,,
}
where we use the standard notation $\phi = h_{00}$. From this series, we can obtain equivalently,
\eq{ \label{velocity-time-subscript}
\hat{U}_0 = -1 + \frac12 \phi + \frac18 \phi^2 - \frac12 \hat{U}^i \hat{U}^i\,.
}

\para{Linearized Einstein equations} \label{linear-einstein}

Our goal is to compute the correction to the average expansion rate induced by the inhomogeneities affecting the spacetime. As shown in section \ref{linear-correction}, the correction which is linear in the perturbations vanishes when averaged over surfaces of constant density. Hence, to get a non-zero correction in this case, we must expand the expansion rate up to the second order in the metric perturbations. However, for the same reason, it suffices to solve the Einstein equations up to the first order, since the second-order contributions to $h_{\A\B}$ would contribute linearly to the expansion rate, and thus they vanish on average.

The theory of linear cosmological perturbations is well-known. Therefore, we will simply state the results relevant to this paper, and refer the reader to the various reviews of the topic, for instance \cite{Mukhanov:1992, Brandenberger:2003}.

In the absence of perturbations, the various fluids contained in the spacetime have a total density $\rho_0$, a pressure $p_0$, and move with the same velocity, $\hat{u}^0_A = 1$ and $\hat{u}_A^i = 0$. When the spacetime becomes slightly inhomogeneous, the fluids can then move with different spatial velocities relative to one another, as well as relative to the background. Expanding their energy-momentum tensor (\ref{energy-momentum}) up to the first order, we get
\eqa{
&T_{00} = a^2\big( \rho_0 + \delta\rho - \rho_0\,\phi\big)\,, \\[5pt]
&T_{0i} = a^2\big( -(\rho_0 + p_0)\hat{u}_i + p_0 \,h_{0i}\big)\,, \\[5pt]
&T_{ij} = a^2\big( p_0 \,\delta_{ij} + \delta p \,\delta_{ij} + p_0 \,h_{ij}\big)\,,
}
where $\delta \rho$ and $\delta p$ denote the total density and pressure perturbations, and $\hat{u}_i$ represents an average spatial velocity for the fluids, given by
\eq{ \label{ave-fluid-velocity}
\hat{u}_i = \sum_A \frac{\rho_{0A} + p_{0A}}{\rho_0 + p_0}\,\hat{u}_i^A\,.
}

The Einstein equations in the homogeneous background, $G_{\A\B}^{(0)} = \kappa T_{\A\B}^{(0)}$, yield the usual Friedmann equations, which can be written as
\eqa{
&\kappa a^2 \rho_0 = 3\Hu^2\,, \\[4pt] 
&\kappa a^2 p_0 = -(2\dot{\Hu} + \Hu^2)\,, 
}
where $\kappa = 8\pi G_N$, and $G_N$ is Newton's gravitational constant. In order to write the Einstein equations at the first order, it is best to decompose the metric perturbations in terms of their scalar, vector, and tensor components, following the standard procedure,
\eqa{ \label{metric-decomposition}
&h_{00} = \phi\,, \\[5pt] 
&h_{0i} = \p_i \theta + \theta_i\,, \\[5pt] 
&h_{ij} = \delta_{ij} \psi + \p_i\p_j \zeta + \p_i \psi_j + \p_j \psi_i + H_{ij}\,, 
}
where $H_{ij}$ represents the transverse-traceless tensor perturbations, with $H_{ii} = 0$ and $\p_j H_{ij} = 0$, and $\theta_i$ and $\psi_i$ are the vector perturbations, with $\p_i \theta_i = \p_i \psi_i = 0$. None of these perturbations, apart from the tensor ones, are invariant under infinitesimal coordinate transformations. However, by combining them with each other, we can define variables which are invariant at the first order in the coordinate changes. The scalar coordinate-invariant variables are
\eqa{ \label{Phi-def}
 \Phi^{(1)} &= \psi + 2\Hu \theta - \Hu \dot{\zeta}\,, \\[4pt]
 \Phi^{(2)} &= \phi - 2\Hu\theta + \Hu \dot{\zeta} - 2\dot{\theta} + \ddot{\zeta}\,,
}
whereas the vector one is
\eq{
 \Phi_i = \theta_i - \dot{\psi}_i\,.
}

It is also worth decomposing the average spatial velocity of the fluids (\ref{ave-fluid-velocity}) in terms of its scalar and vector components,
\eq{
\hat{u}_i = \p_i v + v_i\,,
}
where $v_i$ is the vector component, with $\p_i v_i = 0$. Since the various perfect fluids contained in the spacetime do not generate anisotropic stress at the first order, the linearized Einstein equations $G_{\A\B}^{(1)} = \kappa T_{\A\B}^{(1)}$ lead to a relation between the scalar perturbations,
\eq{ \label{constraint-scalar}
\ddot{\zeta} + 2\Hu \dot{\zeta} - 2(\dot{\theta} + 2\Hu\theta) = \psi - \phi, 
}
which implies that $\Phi^{(1)} = \Phi^{(2)} = \Phi$. In other words, since the two scalar coordinate-invariant variables are equal to each other at the first order, we will denote both of them as $\Phi$. The linearized Einstein equations also imply
\begin{align}
\label{einstein-density} 
\frac{\delta \rho}{\rho_0} &= \phi + \frac{1}{\Hu}\,\dot{\psi} - \frac{1}{3\Hu^2}\, \cov^2\Phi\,,\\[5pt]
\label{einstein-velocity} 
v &= \frac{1}{2(\Hu^2 - \dot{\Hu})} \,(\dot{\psi} + \Hu \phi)\,,\\[4pt]
\label{einstein-velocity-vector}
v_i &= \frac{1}{4(\Hu^2 - \dot{\Hu})}\,\cov^2\Phi_i \,,
\end{align}
where $\cov^2 = \p_i\p_i$ is the Laplacian for the background spatial coordinates. Furthermore, it is customary to neglect the vector components of the metric perturbations, as they always decay with time at the first order, whatever fluids are present in the spacetime. Under special circumstances, these perturbations can nonetheless contribute significantly to the average expansion rate, which is why we choose not to neglect them in this paper. 

\para{Linear-order correction} \label{linear-correction}

As a preliminary exercise, let's compute the expansion rate up to the first order in the metric perturbations. In doing so, we will not take a spatial average, nor will we choose a specific gauge for the perturbations, contrary to what we will do in the rest of the paper. 

Only the scalar perturbations contribute to the expansion rate at the first order. In particular, denoting the scalar component of the spatial velocity of the observed particles as $V$, we have\footnote{Here, $V$ is not itself a velocity, but the scalar component of a velocity. As such, it is not dimensionless, but has units of inverse time.}
\eqa{
\p_i \hat{U}_i &= \cov^2 V \,, \\[4pt]
\p_i \hat{U}^i &= \p_i\big(\hat{U}_i - h_{0i}\big) = \cov^2\big(V - \theta\big)\,.
}

Using (\ref{expansion-rate-series}), (\ref{obs-velocity-time}), (\ref{Phi-def}), and (\ref{einstein-density}), the expansion rate is then, up to the first order,
\eqa{
\hspace{-3pt}\frac{H}{H_0} &= 1 + \frac12 \phi + \frac{1}{2\Hu}\dot{\psi} + \frac{1}{6\Hu}\cov^2\big( \dot{\zeta} - 2\theta + 2V \big) \\[4pt]
&= 1 + \frac{1}{2\rho_0}\,\delta\rho + \frac{1}{6\Hu^2}\cov^2\Psi\,, 
}
where $\Psi$ is defined as
\eq{ \label{psi-definition}
\Psi = \psi + 2\Hu\, V\,.
}

One can verify that the variable $\Psi$ is coordinate-invariant at the first order, under an infinitesimal coordinate transformation. However, the whole linear correction to the expansion rate, which also involves the fluid density perturbation, is not itself coordinate-invariant. Although the value of the full expansion rate $H$ at a fixed point in spacetime is coordinate-invariant, we do not expect its perturbation to be so, since its background value is not constant.

\para{Gauge issue and clock choice} \label{gauge-issue}

The proper way to address the gauge dependence of perturbations in general relativity is to choose coordinates which bear a clear physical interpretation. For instance, in most of this paper, we choose to evaluate the expansion rate over surfaces where the total density of the fluids is constant. In general, these surfaces differ from the ones where the background time $\eta$ is constant. Therefore, one must be particularly careful when averaging the expansion rate over such surfaces, given that its background value $H_0(\eta)$, which depends on the background time and not on the total fluid density, varies from point to point on the surface.

The simplest way around this problem is to choose coordinates for which the surfaces of constant fluid density and background time precisely coincide. This happens in the uniform-density gauge, when $\delta\rho = 0$. The total fluid density is then equal to its background value, which provides a clear physical interpretation for the background time,
\eq{
\rho = \rho_0(\eta)\,.
}
In that gauge, the expansion rate becomes
\eq{
\frac{H}{H_0} = 1 + \frac{1}{6\Hu^2}\,\cov^2\Psi\,.
}

Taking the average over a surface of constant total density (and of constant background time $\eta$), we get
\eq{ \label{linear-correction-vanishes}
\overline{H} \,=\, H_0 + \frac{1}{6\Hu^2} \frac{\int \dd[3]{x} \cov^2\Psi}{\int \dd[3]{x}} \,=\, H_0\,.
}

Thanks to our choice of coordinates, the background expansion rate can indeed be taken out of the integral, as it remains constant on the surface over which we are integrating. By Gauss's theorem, the integral of the perturbation reduces to a surface term, which can be neglected assuming the integration volume is large enough, and the perturbation obeys periodic boundary conditions. Hence, the linear correction to the expansion rate does indeed vanish when averaged over surfaces of constant total density, insofar as we can neglect terms at the boundary, as we will do systematically in this paper.

\section{Average at Constant Total Density} \label{section-ave-total}

\para{Average expansion rate} \label{subsection-average-series}

In this section, we compute the average value of the expansion rate over surfaces where the total fluid density is constant. Later in this paper, we will also compute its average value over other surfaces. 

For this reason, let's first consider the more general problem of averaging a scalar function $F$ over some slicing of spacetime. This function can be written in terms of its value $F_0$ in the background, plus a small perturbation\footnote{In this paper, the perturbation of a function is usually defined by factoring out its background value, as we have done here for $\delta F$. By construction, $\delta F$ is thus dimensionless, rather than having the same units as $F$. The same holds, in particular, for the perturbation $\delta H$ affecting the average expansion rate, as defined in (\ref{def-dH}).},
\eq{ \label{def-F}
F(\eta, \bm{x}) = F_0(\eta)\big( 1 + \delta F (\eta, \bm{x})\big)\,.
}

We want to compute the average of that function up to the second order in the perturbations, over a three-dimensional space-like surface. We assume that the latter has an induced metric $\gamma_{ij}$, with a determinant $\gamma$. In the absence of perturbations, it reduces to the surface of constant background time, with $\gamma_{ij}^{(0)} = a^2\,\delta_{ij}$. Hence, when metric perturbations are included, we get
\eq{ \label{def-gamma}
\sqrt{\gamma} = a^3(\eta)\big(1 + \delta\gamma\big)\,,
}
where $\delta\gamma$ represents the effect of the perturbations. The average value of the function $F$ over this surface is
\eq{ 
\overline{F} = \frac{\int \dd[3]{x}\sqrt{\gamma}\,F}{\int \dd[3]{x}\sqrt{\gamma}}\,.
}

Expanding up to the second order in the perturbations, using (\ref{def-F}) and (\ref{def-gamma}), we get
\eq{ \label{average-F}
\frac{\overline{F}}{F_0} = 1 + \expval{\delta F} + \expval{\delta F\,\delta\gamma} - \expval{\delta F}\expval{\delta\gamma}\,,
}
where the brackets represent the average over surfaces of constant time in the homogeneous background,
\eq{ \label{average-background}
\expval{F} =  \frac{\int \dd[3]{x} F}{\int \dd[3]{x}}\,.
}

It is worth noticing that the induced metric of the slicing contributes to the average value only through products with $\delta F$. Hence, at the second order, it is sufficient to express the induced metric up to the first order in the metric perturbations. 

Let's now deal with the specific case of averaging the expansion rate over surfaces of constant total fluid density. These coincide with the surfaces of constant background time in the uniform-density gauge. Thus, in that gauge, they have an induced metric $\gamma_{ij} = g_{ij} = a^2(\delta_{ij} + h_{ij})$, which implies that
\eq{
\delta \gamma = \frac12 h_{ii} = \frac12 \cov^2 \zeta + \frac32 \psi\,.
}

The linear correction to the expansion rate has been computed in section \ref{linear-correction}. Therefore, up to the second order, it takes the form
\eq{
\frac{H}{H_0} = 1 + \frac{1}{6\Hu^2} \cov^2\Psi + H_2\,,
}
where $H_2$ represents the corrections which are quadratic in the perturbations. Taking the average over surfaces of constant total density, we get
\eq{ \label{def-dH}
\frac{\overline{H}}{H_0} = 1 + \expval{\delta H}\,,
}
where $\delta H$ denotes the effective correction to the expansion rate, which includes in particular the correction from the induced metric of the slicing. Using (\ref{average-F}), it is given by
\eq{ \label{effective-correction}
\delta H = H_2 + \frac{1}{12\Hu^2}\big( \cov^2 \zeta + 3\psi\big) \cov^2 \Psi\,.
}

In this paper, we will assume that the integration volume is large enough, and that perturbations obey periodic conditions at its boundary, so that boundary terms can be neglected. In that case, spatial divergences such as $\p_i F_i$ contained in the correction $\delta H$ vanish when averaged over space, in $\expval{\delta H}$, as they amount to boundary terms by Gauss's theorem. Hence, in order to avoid keeping track of unnecessary terms, we will systematically neglect spatial divergences as soon as they appear in $\delta H$. In doing so, we can simplify expressions by moving around spatial derivatives, for instance
\eqa{
\p_i \phi\, \p_i \psi &= -\phi \cov^2\psi + \p_i (\phi\, \p_i \psi)\\[4pt]
&= -\phi \cov^2\psi + \text{boundary term}\,.
}

\para{Uniform-density gauge}

Since the principle of relativity stipulates that a particular choice of coordinates should not have any impact on the physics, we are free to impose  four additional constraints on the metric perturbations, two for their scalar components, and two for their vector components. In this paper, we choose to use the uniform-density gauge, which we define as
\eqa{
\delta\rho &= 0\,, \\[4pt]
\theta &= 0\,, \\[4pt]
\theta_i &= 0\,.
}
Under these circumstances, the metric becomes
\eqa{
&h_{00} = \phi\,, \\[4pt] 
&h_{0i} = 0\,, \\[4pt] 
&h_{ij} = \delta_{ij} \psi + \p_i\p_j \zeta + \p_i \psi_j + \p_j \psi_i + H_{ij}\,.
}

Our gauge choice regarding the total fluid density, $\delta\rho = 0$, imposes a constraint on the metric perturbations, which can be derived from the Einstein equation $G_{00} = \kappa T_{00}$. At the first order, it is given by (\ref{einstein-density}). Expanding up to the second order, it becomes
\eq{ \label{constraint-phi}
\phi = -\frac{1}{\Hu}\dot{\psi} + \frac{1}{3\Hu^2}\cov^2\Phi + \phi^{(2)}\,,
}
where $\phi^{(2)}$ represents the additional terms which are quadratic in the perturbations. Using (\ref{energy-momentum}) and (\ref{velocity-time-subscript}), the time-time component of the energy-momentum tensor of the fluids is
\eq{
T_{00} = a^2\rho_0\left( 1 - \phi + \sum_A \frac{\rho_{0A} + p_{0A}}{\rho_0} \,\hat{u}_i^A \hat{u}_i^A\right)\,.
}

Therefore, the terms in the constraint on $\phi$ coming from the second-order Einstein equation $G_{00}^{(2)} = \kappa T_{00}^{(2)}$ are
\eq{ \label{phi-second-order}
\phi^{(2)} = -\frac{1}{3\Hu^2} G_{00}^{(2)} + \sum_A \frac{\rho_{0A} + p_{0A}}{\rho_0} \,\hat{u}_i^A \hat{u}_i^A\,.
}

The first term on the right-hand side of this equation represents the second-order perturbation of the Einstein tensor. For coordinates for which $\theta = 0$ and $\theta_i=0$, a long but straightforward calculation yields\footnote{For future reference, we have not used either the constraint $\delta\rho = 0$ or the Einstein equations when deriving this equation. It remains valid in any gauge for which $h_{0i} = 0$.}
\eqa{ \label{einstein-second-order}
G_{00}^{(2)} &= - \frac18 \dot{H}_{ij} \dot{H}_{ij} + \frac18 H_{ij} \cov^2 H_{ij} - \Hu \,h_{ij} \dot{h}_{ij} \\[3pt]
&\hspace{15pt} + \frac14 \Phi_i \cov^2 \Phi_i + \phi\cov^2\psi + \frac54 \psi \cov^2\psi  \\[3pt]
&\hspace{15pt} + \frac12 \psi \cov^4 \zeta + \frac34 \dot{\psi}^2 + \frac12 \dot{\psi}\cov^2\dot{\zeta}\,,
}
where we used the fact that $\Phi_i = -\dot{\psi}_i$ when $\theta_i = 0$. This equation is true up to a spatial divergence, which has been neglected as it reduces to a boundary term when averaged over space. The second term on the right-hand side of (\ref{phi-second-order}) resembles a kinetic energy, associated with the spatial motion of the fluids relative to the background. As it turns out, however, it is best to consider the spatial motion of the fluids relative to the particles used to measure the expansion rate. The velocity of each fluid relative to these observed particle can be defined as
\eq{
\hat{U}_i^A = \hat{u}_i^A - \hat{U}_i\,.
}
where $\hat{U}_i$ denotes the spatial velocity of the observed particles. It is worth decomposing these velocities in terms of their scalar and vector components,
\eqa{ \label{velocities-decomposition}
&\hat{U}_i^A = \p_i V^A + V_i^A\,, \\[4pt]
&\hat{U}_i = \p_i V + V_i\,,
} 
where $V_i^A$ and $V_i$ denote the vector components of the velocities, with $\p_i V_i^A = \p_i V_i = 0$. In terms of the average spatial velocity of the fluids defined in (\ref{ave-fluid-velocity}), we get
\eq{
\sum_A \frac{\rho_{0A} + p_{0A}}{\rho_0} \,\hat{U}_i^A = \frac{2(\Hu^2 - \dot{\Hu})}{3\Hu^2}\,\big( \hat{u}_i - U_i \big)\,,
}
which implies that
\eq{
 \frac12 \sum_A \frac{\rho_{0A} + p_{0A}}{\rho_0} \,\hat{u}_i^A \hat{u}_i^A \,=\,  \frac{\Hu^2 - \dot{\Hu}}{3\Hu^2}\big( 2\,\hat{u}_i \hat{U}_i - \hat{U}_i \hat{U}_i\big)\,+\, \delta E \,,\,\,\,
}
where $\delta E$ represents the kinetic energy of the fluids, relative to the observed particles,
\eq{ \label{kinetic-relative}
\delta E \,=\, \frac12 \sum_A \frac{\rho_{0A} + p_{0A}}{\rho_0} \,\,\hat{U}_i^A \hat{U}_i^A\,.
}
Using (\ref{einstein-velocity}), (\ref{einstein-velocity-vector}), and (\ref{constraint-phi}), we get
\eqa{ \label{kinetic-second-order}
\hspace{-5pt}\frac12 \sum_A \frac{\rho_{0A} + p_{0A}}{\rho_0} \,\hat{u}_i^A \hat{u}_i^A \,&=\, \frac{1}{6\Hu^2} \Phi_i \cov^2 V_i - \frac{1}{9\Hu^3} \Phi\cov^4 V  
\\[3pt] &\hspace{15pt} +\, \frac{\dot{\Hu} - \Hu^2}{3\Hu^2} \hat{U}_i \hat{U}_i \,+\, \delta E \,.
}

\para{Leading-order correction}

We are now ready to compute the effective correction to the average expansion rate. Using (\ref{expansion-rate-series}) and (\ref{obs-velocity-time}), the second-order terms contributing to the expansion rate are
\eqa{
H_2 &= -\frac{1}{6\Hu}h_{ij}\,\dot{h}_{ij} + \frac12 \hat{U}_i \hat{U}_i + \frac{1}{3\Hu}\hat{U}_i \dot{\hat{U}}_i  \\[3pt]  
&\,\hspace{10pt}\, + \frac12 \phi^{(2)} + \frac38 \phi^2 + \frac{1}{4\Hu} \phi\dot{\psi} + \frac{1}{12\Hu}\phi\cov^2\dot{\zeta} \\[3pt]
&\,\hspace{10pt}\,+ \frac{1}{6\Hu} \phi \cov^2 V - \frac{1}{6\Hu} \big( \cov^2\zeta + 3\psi\big)\cov^2 V\,. 
}

The actual correction to the average expansion rate is given by this bare correction, plus the additional terms from the induced metric of the slicing, as stated in (\ref{effective-correction}). At leading order, it can be decomposed in terms of the contributions from the scalar, vector, and tensor perturbations,
\eq{
\delta H = \delta H_S + \delta H_V + \delta H_T\,.
}

Such a decomposition no longer holds at higher orders, where the perturbations of each type mix with one another. At this point, we can make use of the constraints on the metric perturbations from the uniform-density gauge, namely (\ref{phi-second-order}), (\ref{einstein-second-order}), and (\ref{kinetic-second-order}) for $\phi^{(2)}$, as well as the linear constraints
\eqa{
\phi &= -\frac{1}{\Hu}\dot{\psi} + \frac{1}{3\Hu^2}\cov^2\Phi\,, \\[4pt]
\dot{\zeta} &= \frac{1}{\Hu}\big( \psi - \Phi\big) \,.
}

Moreover, using (\ref{psi-definition}) to rewrite $\psi$ in terms of $\Psi$, the scalar, vector, and tensor contributions to the average expansion rate are then respectively
\eqa{ \label{scalar-correction}
\delta H_S &= \frac{1}{24\Hu^2}\Psi\cov^2\Psi - \frac{1}{36\Hu^4} \Psi\cov^2\Phi  - \frac{1}{6\Hu}\Psi\cov^2 V  \\[3pt]
&\,\hspace{10pt}\, - \frac{1}{6\Hu^2}\dot{\Psi} \cov^2 V + \frac{1}{72\Hu^4} \Phi\cov^4\Phi + \delta E_S \,,
}
\vspace{-10pt}
\eqa{ \label{vector-correction}
\delta H_V &= -\frac{1}{24\Hu^2} \,\Phi_i\cov^2\Phi_i + \frac{1}{6\Hu^2} \,\Phi_i \cov^2 V_i \\[3pt]
&\,\hspace{10pt}\, + \frac{\Hu^2 + 2\dot{\Hu}}{6\Hu^2}\, V_i V_i + \frac{1}{3\Hu}\, V_i \dot{V}_i + \delta E_V\,,
}
\vspace{-10pt}
\eq{ \label{tensor-correction}
\delta H_T = \frac{1}{48\Hu^2}\dot{H}_{ij} \dot{H}_{ij} - \frac{1}{48\Hu^2} H_{ij} \cov^2 H_{ij} \,,
}
where $\delta E_S$ and $\delta E_V$ denote the kinetic energies of the fluids, as defined in (\ref{kinetic-relative}), for the scalar and vector components of their relative velocities.

These equations represent the most important result of this paper. Although they have been derived in the uniform-density gauge, they have been written in a way which makes them almost coordinate-invariant. Indeed, most of the variables featured in these equations ($\Phi$, $\Psi$, $\Phi_i$, $V_i$, $H_{ij}$, and $\delta E$) are invariant at the first order under infinitesimal coordinate transformations. Only the scalar component of the velocity of the observed particles $V$ is not coordinate-invariant. Hence, when other gauges are used, this velocity should be replaced by the single coordinate-invariant variable which can be constructed from the scalar perturbations, and which reduces to $V$ in the uniform-density gauge, that is,
\eq{
V^{(\text{gi})} = V + \frac{\delta \rho}{\dot{\rho}_0}\,.
}

Moreover, the author of this paper has verified that the same leading-order correction to the average expansion rate can be obtained by doing the calculation in two other gauges frequently used in cosmology, namely the Poisson gauge and the synchronous gauge. The calculation is much less straightforward in these cases, since the surface of constant density over which we are integrating does not coincide with the surface of constant background time.

The effective correction $\expval{\delta H}$ is expected to be the same whatever coordinates are used for the calculation only because it is properly defined, and bears a clear physical interpretation. This is not true, in particular, for the bare correction to the expansion rate $H_2$, which is indeed different in different gauges. One should therefore avoid reaching any conclusion based solely on a calculation of $H_2$. 

In the long-wavelength limit, when the spatial derivatives of the perturbations can be neglected, the scalar and tensor corrections vanish entirely\footnote{The case of the tensor correction requires some attention. In the long-wavelength limit, the tensor perturbations satisfy the equation
\eqn{
\ddot{H}_{ij} + 2\Hu \dot{H}_{ij} = 0 \,,
}
which implies that
\eqn{
\dot{H}_{ij} = \frac{C_{ij}}{a^2}\,,
}
where $C_{ij}$ is the integration constant. The time derivative of the tensor perturbations decays with the spacetime expansion in the long-wavelength limit. Its contribution to the average expansion rate is thus negligible in most circumstances relevant to cosmology (for instance, when $w<1$ in the case of a spacetime containing a single fluid). Moreover, the long-wavelength effect discussed in \cite{Tsamis:1996a, Tsamis:1996b} is of higher order, and therefore does not contradict our result.}. Indeed, as shown in \cite{Geshnizjani:2002}, the expansion rate of a spacetime containing a single scalar field reduces to its background value in the long-wavelength limit, when averaged over surfaces of constant field. The spacetime considered in this paper contains more than one fluid, but its expansion rate is averaged over surfaces where the total density of the fluids is constant. Our analysis confirms that, over such a surface and in the long-wavelength limit, the various fluids have the same effect on the expansion rate as if they formed a single field. 

On the other hand, the correction induced by the vector perturbations does not necessarily vanish in the long-wavelength limit, when more than one fluid is present in the spacetime. However, since the vector metric perturbations decay with time in most situations relevant to cosmology, there can be a non-zero correction in that limit only when the respective flows of the various fluids are not all irrotational, although they are constrained to be so on average.

Finally, the scalar and vector corrections can contribute both positively and negatively to the average expansion rate, depending on the circumstances. The tensor correction, however, is necessarily positive; it is in fact proportional to the energy density contained in the tensor metric perturbations.

\para{Back-reaction equation of state}\label{back-reaction-eq-state}

In many papers \cite{Buchert:1999, Buchert:2001}, the back-reaction of inhomogeneities is considered in terms of its effects on the averaged Einstein equations. Indeed, when the Einstein equations are averaged over some three-dimensional space-like surface in the presence of perturbations, the resulting equations do not reduce to their analogues in a purely homogeneous spacetime, but contain additional terms.

From the analysis presented in this paper, we can derive one such ``averaged'' equation, namely the analogue of the first Friedmann equation. In the uniform-density gauge, the total fluid density is equal to its background value, which in turns satisfies the background version of the first Friedmann equation,
\eq{
\kappa \rho = \kappa \rho_0 = 3H_0^2\,.
}
Using (\ref{def-dH}), we get at the leading order,
\eq{
\kappa (\rho + \rho_\text{br}) = 3\overline{H}^2\,,
}
where $\rho_\text{br}$ represents an effective density induced by the back-reaction of perturbations, 
\eq{ \label{br-density}
\rho_\text{br} = 2\rho\, \expval{\delta H}\,.
}

Hence, the first Friedmann equation remains valid on average, but with an additional term contributing to the total density. Let's now consider the particular case when, at a specific moment in the history of a spacetime, the effective correction to its average expansion rate varies as $\expval{\delta H} \sim a^n$. Moreover, at that same epoch, the spacetime is dominated by a fluid with a constant equation of state $w$. The total fluid density then varies approximately as
\eq{
\rho \sim a^{-3(1+w)}\,.
}

Under such circumstances, the density induced by the perturbations evolves following a similar power law, but with a different equation of state,
\eq{
\rho_\text{br} \sim a^{-3(1+w_\text{br})}\,,
}
where $w_\text{br}$ is given by
\eq{
w_\text{br} = w - \frac13 n\,.
}

\section{Applications}\label{section-applications}

\para{Average-velocity observed particles}

The purpose of this section is to apply the general formulas we just derived to specific cases, for which the time-dependence of the average expansion rate can be explicitly computed. Let's first consider the case when the particles used to measure the expansion rate move at the average velocity of the fluids (\ref{ave-fluid-velocity}),
\eq{
\hat{U}_i = \hat{u}_i\,.
}

This situation can occur in particular if the fluids, including the particles used to track the spacetime expansion, all move at the same velocity because of their strong interactions with one another. In that case, the kinetic energy relative to their average motion vanishes, $\delta E = 0$. The various fluids then form what can be called a ``multi-component single fluid'' \cite{Andersson:2006, Rezzolla:2013}. This situation also occurs, \textit{a fortiori}, if the spacetime contains only one fluid.

With $V=v$, we can combine (\ref{einstein-velocity}) and (\ref{psi-definition}) to rewrite $\Psi$ in terms of the scalar metric perturbation,
\eq{
\Psi = \psi + \frac{\Hu}{\Hu^2 - \dot{\Hu}}\big(\dot{\psi} + \Hu\phi\big)\,,
} 
that is, using (\ref{Phi-def}) and (\ref{constraint-scalar}),
\eq{ \label{psi-def-2}
\Psi = \frac{\Hu}{\Hu^2 - \dot{\Hu}}\, \dot{\Phi} + \frac{2\Hu^2 -\dot{\Hu}}{\Hu^2 - \dot{\Hu}}\,\Phi\,.
}

This expression for $\Psi$ is valid in any gauge, as long as the observed particles move at the average velocity of the fluids. In the uniform-density gauge, the latter is given by
\eq{
v = \frac{1}{6\Hu(\Hu^2 - \dot{\Hu})} \cov^2 \Phi\,.
}
The scalar correction (\ref{scalar-correction}) then becomes
\eqa{ \label{scalar-correction-2}
\delta H_S &= \frac{1}{24\Hu^2} \Psi \cov^2 \Psi + \frac{1}{72\Hu^4}\Phi\cov^4\Phi \\[3pt]
&\, - \frac{1}{36\Hu^4}\left( \frac{2\Hu^2 - \dot{\Hu}}{\Hu^2 - \dot{\Hu}}\,\Psi + \frac{\Hu}{\Hu^2 - \dot{\Hu}}\,\dot{\Psi} \right) \cov^4 \Phi + \delta E_S\,. 
}

Similarly, for the vector correction, we can take $V_i = v_i$. The linearized Einstein equations imply that the coordinate-invariant vector perturbation must satisfy
\eq{ \label{linearized-eq-vector}
\dot{\Phi}_i + 2\Hu\,\Phi_i = 0\,.
}

Combining this equation with (\ref{einstein-velocity-vector}) and (\ref{vector-correction}), the vector correction then becomes
\eqa{ \label{vector-correction-2}
\delta H_V &= -\frac{1}{24\Hu^2} \Phi_i \cov^2 \Phi_i \\[3pt]
&\hspace{10pt} + \frac{\Hu^4 - 7\Hu^2\dot{\Hu} + 2\dot{\Hu}^2 + 2\Hu\ddot{\Hu}}{96\Hu^2(\Hu^2 - \dot{\Hu})^3}\,\Phi_i \cov^4\Phi_i\, + \,\delta E_V\,.
}

\para{Initial power spectrum}

The corrections we just derived can be expressed in terms of the initial power spectrum of the metric perturbations. Since the background is spatially flat, we can express the scalar metric perturbation in terms of its Fourier modes, with wavenumbers $\bm{k}$,
\eq{
\Phi_{\bm{k}}(\eta) = \int \dd[3]{x}\, e^{i\bm{k}\cdot\bm{x}}\, \Phi(\bm{x})\,,
}

To simplify the calculation, we will consider $\eta = 0$ as our initial time. At that time, the Fourier modes of the scalar metric perturbation take finite values $\Phi_{\bm{k}}(0)$, and so their values at later time are
\eq{ \label{def-f}
\Phi_{\bm{k}}(\eta) = \Phi_{\bm{k}}(0)\,f_k(\eta)\,,
}
where $f_k(0) = 1$ by construction. The scalar correction involves products of perturbations to which are applied certain differential operators. In terms of their Fourier modes, we thus get
\eq{
\delta H_S = \int \frac{\dd[3]{k}}{(2\pi)^3} \frac{\dd[3]{k'}}{(2\pi)^3} \,e^{i(\bm{k}-\bm{k}')\cdot\bm{x}} \,\Phi_{\bm{k}}(0)\,\Phi_{\bm{k}'}^*(0) \,T_{kk'}(\eta)\,,
}
where $T$ is a function related to $f$. Taking the spatial average, and applying Parseval's theorem, we get
\eq{ \label{def-T-S}
\expval{\delta H_S} = \int \frac{\dd{k}}{k}\, P_S(k) \, T_S(k,\eta)\,,
}
where $T_S = T_{kk}(\eta)$, and $P_S$ is the initial power spectrum of the scalar metric perturbation,
\eq{
P_S(k) = \frac{k^3}{2\pi^2\,V}\,\abs{\Phi_{\bm{k}}(0)}^2\,,
}
where $V = \int \dd[3]{x}$ denotes the integration volume. This factor can be absorbed in the initial value of the perturbation, by an appropriate redefinition of the Fourier transform. A similar result can be derived for the tensor metric perturbations,
\eq{ \label{def-T-T}
\expval{\delta H_T} = \int \frac{\dd{k}}{k}\, P_T(k) \, T_T(k,\eta)\,,
}
where $P_T$ is the initial power spectrum of the tensor metric perturbations,
\eq{
P_T(k) = \frac{k^3}{2\pi^2\,V}\,H_{ij}(\bm{k},0)H_{ij}^*(\bm{k}, 0)\,.
}

In this paper, we will assume that the initial power spectra of the scalar and tensor perturbations are approximately scale-invariant. Such a spectrum is indeed produced if the epoch under consideration is preceded, for instance, by a phase of inflationary expansion.

\para{Cold matter}

In the following sections, we will consider the case of a spacetime containing a single fluid, with a constant equation of state parameter $w$. The background expansion rate in comoving coordinates is then
\eq{
\Hu = \frac{\alpha}{\eta} = \frac{2}{(1+3w)\,\eta}\,.
}

The background scale factor varies with conformal time as $a \sim \eta^\alpha$. Furthermore, since the pressure perturbation is purely adiabatic in this case, the scalar metric perturbation satisfies the equation\footnote{For instance, see the equation (5.22) in \cite{Mukhanov:1992}.}
\eq{ \label{eq-scalar-perturbation}
\ddot{\Phi} + \frac{2(\alpha+1)}{\eta}\,\dot{\Phi} - \frac{2-\alpha}{3\alpha} \,\cov^2 \Phi = 0\,.
}

Let's first consider the case of pressureless ``cold'' matter, for which $w=0$ and $\alpha = 2$. Since the term with a Laplacian involved in the previous equation vanishes in that case, there is no need to take a Fourier transform. Integrating yields
\eq{
\Phi(\eta, \bm{x}) = A(\bm{x}) + \frac{B(\bm{x})}{\eta^5}\,,
}
where $A(\bm{x})$ and $B(\bm{x})$ are functions of the spatial coordinates which can be fixed by the initial conditions. For the perturbation to be initially finite, at $\eta = 0$, we must take $B = 0$. We can also ignore this solution on the basis that it decays with the spacetime expansion. In any case, since the scalar perturbation remains constant, it follows from (\ref{psi-def-2}) that
\eq{
\Psi = \frac{5}{3} \Phi\,.
} 

Using (\ref{scalar-correction-2}), the scalar correction to the average expansion rate of a spacetime containing cold matter is thus
\eq{
\expval{\delta H_S} = -\frac{25}{864}\,\eta^2 \expval{(\cov \Phi)^2} - \frac{41}{10368}\,\eta^4 \expval{(\cov^2\Phi)^2}\,.
}

The correction is negative at all times. The effective density (\ref{br-density}) induced by this correction in the first Friedmann equation is therefore also negative, $\rho_\text{br} < 0$. At early times, the correction varies as $\expval{\delta H_S} \sim \eta^2$, which corresponds to $n = 1$ and $w_\text{br} = -\frac13$. At later times, the second term in the previous equation eventually becomes dominant compared to the first term, and then $n=2$ and $w_\text{br} = -\frac23$. 

This simple example illustrates the general effect of perturbations on the average expansion rate. At early times, the back-reaction behaves as a fluid with an equation of state $w_\text{br} = -\frac13$. As such, it tends to mimic the effects of constant spatial curvature in the homogeneous background. This confirms the analysis of previous work \cite{Geshnizjani:2005, Blachier:2023}, which have also come to the same conclusion.

However, it should be stressed that the back-reaction equation of state is not constant. Hence, the effect of perturbations on the average expansion rate cannot be reduced to that of constant spatial curvature. For a spacetime containing cold matter, the equation of state becomes more negative with time, eventually reaching $w_\text{br} = -\frac23$, causing a deceleration of the expansion\footnote{It is indeed a deceleration and not an acceleration, since $\rho_\text{br} < 0$.}. 

\para{Radiation}

Let's now consider the case of a spacetime containing radiation, with $w=\frac13$ and $\alpha = 1$. The scalar metric perturbation then satisfies the equation
\eq{
\ddot{\Phi} + \frac{4}{\eta} \,\dot{\Phi} - \frac13 \cov^2 \Phi = 0\,.
}

Taking the Fourier transform, and rewriting everything in terms of $x=\frac{1}{\sqrt{3}}\, k\eta$, we get
\eq{
f'' + \frac{4}{x} f' + f = 0\,,
}
where a prime denotes a derivative with respect to $x$. The function $f$ was defined in (\ref{def-f}), to factor out the initial value of the perturbation. Fixing the integration constants with $f(0) =1$, we get
\eq{
f = \frac{\Phi_{\bm{k}}}{\Phi_{\bm{k}}(0)} = \frac{3}{x^3} \big( \sin x - x \cos x\big)\,,
}
and so, from (\ref{psi-def-2}),
\eq{
\frac{\Psi_{\bm{k}}}{\Phi_{\bm{k}}(0)} = \frac{3}{2x}\,\sin x\,.
}

Plugging these in (\ref{scalar-correction-2}), we can derive the function $T_S$ defined in (\ref{def-T-S}),
\eqa{
T_S &= \frac{9}{32}\Bigg( x^2 - \frac12 + \frac{2}{x^2} + \left( x - \frac{4}{x}\right) \sin(2x) \\[3pt]&\,\hspace{45pt} + \left(x^2 + \frac92 - \frac{2}{x^2}\right) \cos(2x)\Bigg)\,.
}

This function contains the full time-dependence of the scalar contribution to the average expansion rate, in the presence of radiation. It can be expanded for small and large values of $x$,
\eqa{
T_S &= -\frac{9}{32} \,x^2 \hspace{70pt} (x \ll 1)\,, \\[4pt]
T_S &= \frac{9}{32} \,x^2 \big(1 + \cos(2x)\big) \hspace{20pt} (x \gg 1)\,.
}

As explained earlier, we will assume that the initial power spectrum of the scalar metric perturbation is approximatively scale-invariant, namely that it takes the same value for all wavenumbers below a certain maximum $k_c$. Indeed, the Fourier modes with very large wavenumbers should not contribute significantly to the correction, otherwise the latter would become infinite. These modes correspond to inhomogeneities with very small wavelengths, for which our description of the spacetime energy-momentum in terms of classical, perfect fluids breaks down. 

For such an initial power spectrum, the scalar correction ({\ref{def-T-S}) becomes
\eq{
\expval{\delta H_S} = P_S \int_0^{x_c} \frac{\dd{x}}{x} \,T_S(x)\,,
}
where $x_c = \frac{1}{\sqrt{3}}\,k_c\eta$. Hence, where the function $T_S$ only depends on $k\eta$, as it does when the spacetime contains a single fluid with a constant equation of state, all the time-dependence of the correction can be included in the upper bound of the integral. Integrating,
\eqa{
\expval{\delta H_S} &= \frac{9}{64}\,P_S\bigg( x_c^2 + x_c\sin(2x_c) - \ln(2x_c) 
\\[3pt]&\,\hspace{40pt} - \frac12 \cos(2x_c) + \frac{4\sin(2x_c)}{x_c} - \frac{2}{x_c^2}
\\[3pt]&\,\hspace{40pt} + \frac{2\cos(2x_c)}{x_c^2} + \Ci(2x_c) - \gamma - \frac72\bigg)\,,
}
where $\gamma$ denotes the Euler-Mascheroni constant. Expanding this function at early and late times, for small and large values of $x_c$, we get
\eqa{
\expval{\delta H_S} &= -\frac{3}{64}\,P_S\,(k_c\eta)^2 \hspace{25pt} (k_c\eta \ll 1)\,, \\[4pt]
\expval{\delta H_S} &= \frac{3}{64}\,P_S\,(k_c\eta)^2 \hspace{32pt} (k_c\eta \gg 1)\,.
}

The correction is negative at early times, reaches a minimum value near $k_c\eta \simeq 3.63$, and  then starts growing quadratically with time, becoming positive at $k_c\eta \simeq\, 5.02$. At both early and late times, the correction varies as $\expval{\delta H_S} \sim \eta^2$, which corresponds to $w_\text{br} = -\frac13$. Hence, in a spacetime containing radiation, the back-reaction of the scalar perturbations behaves as constant spatial curvature, whose sign is positive at early times, and negative at late times.

\para{Positive equation of state}

We can generalize the previous calculation by considering a fluid with a general positive equation of state $w>0$. In that case, the equation (\ref{eq-scalar-perturbation}) satisfied by the Fourier modes of the scalar metric perturbation can be written in terms of $x = \sqrt{w}\, k\eta$. Solving this equation, we get
\eq{
f = \frac{\Phi_{\bm{k}}}{\Phi_{\bm{k}}(0)} = 2^{\alpha + \frac12} \,\Gamma\Big(\alpha + \frac32\Big) \frac{J_{\alpha+\frac12}(x)}{x^{\alpha + \frac12}}\,,
}
where we have fixed the integration constants by imposing that $f(0) = 1$. Expanding in series for small and large values of $x$, we get
\eqa{ \label{f-scalar-general}
f &= 1 - \frac{1}{2(2\alpha + 3)}\,x^2 \hspace{100pt} (x \ll 1)\,, \\[4pt]
f &= \frac{2^{\alpha+1}\,\Gamma\Big( \alpha + \frac32\Big)}{\sqrt{\pi}}\,\frac{\cos\left(x - \frac12 \alpha\pi - \frac12\pi\right)}{x^{\alpha+1}} \hspace{16pt} (x \gg 1)\,.
}

Since Bessel functions are not easily integrated, we will focus here on the early and late time limits of the scalar correction, rather than computing its exact expression as we did in the previous section. It turns out that the terms in $\delta H_S$ which dominate in these limits are respectively
\eqa{
\delta H_S &= \frac{1}{24\Hu^2} \,\Psi\cov^2\Psi \hspace{80pt} (k\eta \ll 1)\,,\\[4pt]
\delta H_S &= -\frac{1}{36\Hu^3 (\Hu^2 - \dot{\Hu})}\, \dot{\Psi} \cov^4 \Phi \hspace{30pt} (k\eta \gg 1)\,.
}
Using (\ref{f-scalar-general}), the function $T_S$ defined in (\ref{def-T-S}) becomes
\eqa{
T_S &= -\frac{(2\alpha + 1)^2}{24\alpha^2(\alpha+1)^2} \,(k\eta)^2 \hspace{60pt} (k\eta \ll 1)\,,\\[4pt]
T_S &= \frac{(12\alpha)^\alpha\,\Gamma^2\Big( \alpha + \frac32\Big)}{18\pi\,\alpha^4(\alpha+1)^2(2-\alpha)^\alpha} \,(k\eta)^{4-2\alpha} \hspace{10pt} (k\eta \gg 1)\,.
}

Assuming the initial power spectrum of the scalar metric perturbation is scale-invariant, as described in the previous section, we get at early times
\eqa{
\expval{\delta H_S} &= P_S \int_0^{k_c} \frac{\dd{k}}{k}\,T_S \\[4pt]
&= -\frac{(2\alpha + 1)^2}{48\alpha^2(\alpha+1)^2} \,P_S\,(k_c\eta)^2 \hspace{20pt} (k_c \eta \ll 1)\,.
}

At late times, we must be a little more careful, since the Fourier modes with small wavenumbers still contribute to the overall correction. However, their contribution can be neglected as it  merely amounts to a constant. The main contribution comes from the large wavenumbers,
\eqa{
\expval{\delta H_S} &= \frac{(12\alpha)^\alpha\, \Gamma^2\Big( \alpha + \frac32\Big)}{36\pi\,\alpha^4(\alpha+1)^2(2-\alpha)^{\alpha+1}} \,P_S\,(k_c\eta)^{4-2\alpha} 
\\[6pt]&\,\hspace{140pt} (k_c \eta \gg 1)\,.
}

Taking $\alpha=1$ in the two last equations, we obtain the same early and late time limits as we got in the case of radiation, from the exact expression of the correction. As for radiation, the correction is negative at early times, and eventually becomes positive at late times. Initially, it varies as $\expval{\delta H_S} \sim a^{1 + 3w}$, which corresponds once again to $w_\text{br} = -\frac13$. For any fluid with a constant equation of state, the back-reaction of perturbations has the same effects as constant spatial curvature in the homogeneous background. 

At late times, however, the correction varies instead as $\expval{\delta H_S} \sim a^{6w}$, which corresponds to $w_\text{br} = -w$. If $w>\frac13$, the back-reaction equation of state becomes more negative with time, thus causing an acceleration of the expansion.

\para{Tensor perturbations}

Let's now turn to the correction induced by the tensor perturbations, once again for a spacetime containing a single fluid with a constant equation of state. At the linear order, the tensor metric perturbations obey the equation
\eq{
\ddot{H}_{ij} + \frac{2\alpha}{\eta}\,\dot{H}_{ij} - \cov^2 H_{ij} = 0\,.
}

As in the scalar case, we can express the Fourier modes of the perturbations in terms of a function $f$, by factoring out their initial values,
\eq{
H_{ij}(\bm{k}, \eta) = H_{ij}(\bm{k},0)\,f_k(\eta)\,.
}
The function $f$ must then satisfy the equation
\eq{ \label{f-tensor-eq}
f'' + \frac{2\alpha}{x}\,f' + f = 0\,,
}
where $x = k\eta$, and a prime represents a derivative with respect to $x$. Moreover, combining (\ref{tensor-correction}) and (\ref{def-T-T}), the tensor contribution to the average expansion rate is given by
\eq{ \label{T-T-single-fluid}
T_T = \frac{1}{48\alpha^2} \,x^2\,\big( f'^2 + f^2 \big)\,.
}

This correction is positive under all circumstances, at all times and for all types of fluids, as it represents the energy density of the tensor perturbations. In particular, let's focus on the case when the spacetime contains radiation, for which $w = \frac13$ and $\alpha = 1$. The solution of (\ref{f-tensor-eq}) satisfying the initial condition $f(0) = 1$ is then
\eq{
f = \frac{\sin x}{x}\,.
}
Plugging this function in the previous equation, we get
\eq{ 
T_T = \frac{1}{48}\left( 1 + \frac{1}{2x^2} - \frac{\sin(2x)}{x} - \frac{\cos(2x)}{2x^2}\right)\,.
}

Assuming the initial power spectrum of the tensor perturbations is approximatively scale-invariant, the correction they induce becomes
\eq{ \label{tensor-correction-scale-invariant}
\expval{\delta H_T} = P_T \int_0^{x_c} \frac{\dd{x}}{x} \,T_T(x)\,,
}
where $x_c = k_c \eta$. For pure radiation, we get
\eqa{
\expval{\delta H_T} &= \frac{1}{48} P_T \bigg( \ln(2x_c) + \frac{\sin(2x_c)}{2x_c} - \frac{1}{4x_c^2} 
\\[3pt]&\,\hspace{40pt} + \frac{\cos(2x_c)}{4x_c^2}  - \Ci(2x_c) + \gamma - \frac12 \bigg)\,,
}
that is, at early and late times,
\eqa{
\expval{\delta H_T} &= \frac{1}{96}\,P_T\,(k_c\eta)^2 \hspace{35pt} (k_c \eta \ll 1)\,, \\[4pt]
\expval{\delta H_T} &= \frac{1}{48}\,P_T\,\ln(k_c\eta) \hspace{30pt} (k_c \eta \gg 1)\,.
}

Initially, the correction grows quadratically with time, and thus behaves as negative spatial curvature. In \cite{Abramo:1997}, the average energy-momentum tensor representing the second-order effects of the tensor perturbations also leads to $w_\text{br} = -\frac13$, but with a different sign for the spatial curvature. This difference might come from the fact that the calculations in this paper rely on the Poisson gauge rather than the uniform-density gauge. At late times, the correction continues to grow, but at a much slower rate, varying only logarithmically with time.

We can generalize the previous calculation by considering a fluid whose equation of state is included in $ -\frac13 < w < 1$. In that case, the solution of (\ref{f-tensor-eq}) is
\eq{
f = 2^{\alpha - \frac12} \,\Gamma\Big(\alpha + \frac12 \Big)\, \frac{J_{\alpha -\frac12}(x)}{x^{\alpha-\frac12}}\,.
}
Expanding in series for small and large arguments, we get
\eqa{
f &= 1 - \frac{1}{2(2\alpha+1)}\,x^2 \hspace{82pt} (x\ll 1)\,, \\[4pt] 
f &= \frac{2^\alpha\,\Gamma\left( \alpha + \frac12 \right)}{\sqrt{\pi}}\,\frac{\cos\left( x - \frac12 \alpha\pi \right)}{x^\alpha} \hspace{35pt} (x\gg 1)\,.
}
In these limits, the function $T_T$ given in (\ref{T-T-single-fluid}) becomes
\eqa{
T_T &= \frac{1}{48\alpha^2}\,x^2 \hspace{82pt} (x \ll 1)\,, \\[4pt]
T_T &= \frac{4^{\alpha}\,\Gamma^2\left( \alpha + \frac12 \right)}{48\pi\,\alpha^2}\,x^{2-2\alpha} \hspace{30pt} (x \gg 1)\,.
}

Plugging in (\ref{tensor-correction-scale-invariant}), the tensor correction is then, at early at late times,
\eqa{
\expval{\delta H_T} &= \frac{1}{96\alpha^2}\,P_T\,(k_c\eta)^2 \hspace{70pt} (k_c\eta \ll 1)\,, 
\\[4pt]
\expval{\delta H_T} &= \frac{4^{\alpha}\,\Gamma^2\left( \alpha + \frac12 \right)}{96\pi\,\alpha^2(1-\alpha)}\,P_T\,(k_c\eta)^{2-2\alpha} \hspace{15pt} (k_c\eta \gg 1)\,,
}

Taking the limit when $w=\frac13$ or $\alpha = 1$, these equations coincide with the ones we obtained for a spacetime containing radiation. Once again, at early times, the correction emulates negative spatial curvature, $w_\text{br} = -\frac13$. 

When computing the late-time limit of the correction, we have neglected the contribution coming from the lower bound of the integral, which amounts to a constant. However, this constant becomes dominant compared to the correction we just obtained whenever the latter decreases with time, that is, when $ -\frac13 < w < \frac13$. In these cases, the tensor perturbations do not significantly alter the average expansion rate at late times, and their effect on the first Friedmann equation resembles that of the fluid contained in the spacetime, $n=0$ and $w_\text{br} = w$. 

On the other hand, when $ \frac13 < w < 1$, the tensor contribution to the average expansion rate continues to grow at late times, although less rapidly than it does initially. The correction we derived is thus dominant compared to the constant contribution from the lower bound of the integral. It behaves as radiation, $w_\text{br} = \frac13$.

\para{Vector perturbations}

As a final example, let's compute the correction induced by the vector perturbations, once again for a spacetime containing a single fluid with a constant equation of state. In that case, integrating (\ref{linearized-eq-vector}) yields
\eq{
\Phi_i = \frac{C_i(\bm{x})}{\eta^{2\alpha}}\,,
}
and so the vector contribution (\ref{vector-correction-2}) to the average expansion rate is then
\eqa{
\expval{\delta H_V} &= \frac{1}{24\alpha^2}\,\expval{(\cov C)^2} \,\eta^{2-4\alpha} \\[3pt]
&\hspace{10pt} + \frac{\alpha + 6}{96\alpha^3 (\alpha+1)^2}\, \expval{(\cov^2 C)^2} \,\eta^{4-4\alpha}\,.
}

For most fluids, the two terms involved in the previous correction both decay as the spacetime expands. However, for fluids with $w>\frac13$, the second term of the correction grows with time, and can thus significantly contribute to the average expansion rate, despite the fast decay of the vector metric perturbations. Moreover, when $w>1$, the first term also contributes to the correction.

At early times, it actually dominates the correction, and induces the same effects in the first Friedmann equation as a stiff fluid, with $w_\text{br} = 1$. At late times, the second term becomes dominant compared to the first one, and grows as $\expval{\delta H_V} \sim a^{6w-2}$, which corresponds to $w_\text{br} = -w + \frac23$. In all these cases when the vector correction cannot be neglected, its contribution to the average expansion rate is always positive. Hence, when the fluid equation of state is large enough, more precisely when $w>1$, the back-reaction of the vector perturbations can cause an acceleration of the expansion at late times.

\section{Average for a Different Fluid Property} \label{section-ave-partial}

\para{Another fluid property as cosmic clock}

So far in this paper, we have considered the average value of the expansion rate over a surface where the total density of the fluids is constant. However, this surface might not be observationally relevant. We might want to average the expansion rate instead over other surfaces, for instance one where the density of a single fluid component is constant, rather than that of all the fluids contained in the spacetime. We might also want to use a different physical property of the fluids as our clock, for instance their pressure or temperature.

The purpose of this section is to address this issue, and to compute the average expansion rate over a surface where some scalar property of the fluids, say $\chi$, is constant. To do so, one might consider repeating the long calculation carried out in section \ref{section-ave-total} all over again, but in a different gauge.  However, once the average expansion rate over a given surface has been obtained, its value over a different surface can be related to the former by means of a purely geometrical calculation, which does not rely on the Einstein equations, or on the particular definition of the expansion rate. 

In other words, we will compare the average values of the expansion rate over surfaces of constant total density, and of constant fluid property $\chi$. The latter can be expressed in terms of its value $\chi_0$ in the background, plus a small perturbation,
\eq{
\chi(\eta, \bm{x}) = \chi_0(\eta) + \delta\chi(\eta, \bm{x})\,.
}

In the uniform-density gauge, the total fluid density is equal to its background value, and thus depends only on the background time $\eta$. Similarly, we can define another time $\eta_\chi$, for which an analogous situation holds for the fluid property $\chi$,
\eq{
\chi(\eta, \bm{x}) = \chi_0\big(\eta_\chi(\eta, \bm{x})\big)\,.
}

Since $\chi$ and $\eta_\chi$ are functions of one another by construction, the surfaces where they are respectively constant coincide with each other. In the absence of perturbations, $\eta_\chi$ reduces to the background time. When the spacetime becomes slightly inhomogeneous, it gets modified by a small correction $\delta\eta$,
\eq{
\eta_\chi(\eta, \bm{x}) = \eta - \delta \eta(\eta, \bm{x})\,.
}

Combining the three last equations, and expanding up to the first order in the perturbations, we get 
\eq{ \label{eta-density}
\delta\eta = -\frac{\delta\chi}{\dot{\chi}_0}\,.
}

The perturbation $\delta\eta$ encodes the difference between the surfaces of constant total density, and of constant fluid property $\chi$. Therefore, it is not surprising that the difference between averages over these two surfaces depends essentially on $\delta\eta$, as we will show in the next section.

\para{Average comparison}

The precise definition of the expansion rate, in terms of the velocity of fluid particles, does not matter when it comes to comparing its average values over two surfaces. Hence, in this section, we will consider instead a general scalar function $F$. The latter can be written in terms of its background value $F_0$, plus a small perturbation,
\eq{ \label{def-function-F}
F(\eta, \bm{x}) = F_0(\eta)\big(1 + \delta F(\eta, \bm{x}) \big)\,.
}

We want to compare the average value of this function over a surface of constant fluid property $\chi$, to its value on a surface of constant total fluid density. In order to make such a comparison, the two averages will be calculated at the same moment in the history of the spacetime, more precisely when the cosmic times associated with the integration surfaces become equal to the same numerical value $\eta_0$\footnote{We could have chosen a different ``calibration'' constraint in order to define the moment when the two averages are to be compared. For instance, we could choose different values for the cosmic times, say $\eta=\eta_0$ and $\eta_\chi = \eta_0 + \delta\eta_0$. Here, $\delta\eta_0$ must be constant, otherwise the surface defined by the constraint would not be one where $\chi$ is constant. The only effect of this small difference in the cosmic times would be to shift the perturbation $\delta\eta$ by a small constant amount.}. In that case, the averages are defined as
\eqa{
&\expval{F}_\eta = \frac{\int \dd[3]{x}\, \sqrt{\gamma}\, F\big|_{\eta=\eta_0}}{\int \dd[3]{x}\, \sqrt{\gamma}\big|_{\eta=\eta_0}}\,, \\[5pt]
&\expval{F}_{\eta_\chi} = \frac{\int \dd[3]{x}\, \sqrt{\gamma_\chi}\, F\big|_{\eta_\chi=\eta_0}}{\int \dd[3]{x}\, \sqrt{\gamma_\chi}\big|_{\eta_\chi=\eta_0}}\,,
}
where $\gamma$ and $\gamma_\chi$ denote the determinants of the induced metrics describing each surface. Up to the first order in the perturbations, these induced metrics coincide with each other, and with the spatial part of the background metric, $\gamma_{ij} = (\gamma_\chi)_{ij} = g_{ij}$. The determinants for the two surfaces differ nonetheless from one another, as they are not evaluated at the same points. Following (\ref{def-gamma}), we can define
\eqa{ \label{def-gammas}
&\sqrt{\gamma} = a^3(\eta)\big( 1 + \delta \gamma\big)\,, \\[4pt]
&\sqrt{\gamma_\chi} = a^3(\eta_\chi)\big( 1 + \delta \gamma_\chi\big)\,.
}
With $\eta = \eta_\chi + \delta\eta$, we get at the first order
\eq{ \label{relation-gammas}
\delta \gamma_\chi = \delta \gamma + 3\Hu\,\delta\eta\,.
}

Moreover, using (\ref{def-function-F}) and (\ref{def-gammas}), we can expand the average value of the function $F$ in series. The calculation has already been carried out up to the second order, and the result (\ref{average-F}) can be written equivalently as
\eq{
\expval{F}_\eta = \expval{F|_{\eta=\eta_0}} + \expval{F|_{\eta=\eta_0}\,\delta\gamma} - \expval{F|_{\eta=\eta_0}}\expval{\delta\gamma}\,,
}
where the brackets without any subscript denote a spatial average in the background, as defined in (\ref{average-background}). The previous equation is valid for any function $F$, in particular for the perturbation $\delta F$,
\eq{ \label{ave-series-perturbation}
\expval{\delta F}_\eta = \expval{\delta F|_{\eta=\eta_0}} + \expval{\delta F\,\delta\gamma} - \expval{\delta F}\expval{\delta \gamma}\,.
}

Similarly, with $\eta=\eta_\chi + \delta\eta$, the average of the perturbation over the other surface becomes
\eqa{
\expval{\delta F}_{\eta_\chi} &= \expval{\delta F|_{\eta_\chi=\eta_0}} + \expval{\delta F\,\delta\gamma_\chi} - \expval{\delta F}\expval{\delta \gamma_\chi} \\[3pt]
&= \expval{\delta F|_{\eta=\eta_0}} + \expval{\delta \dot{F} \,\delta\eta} + \expval{\delta F\,\delta\gamma_\chi} - \expval{\delta F}\expval{\delta \gamma_\chi}\,.
}
Combining this equation with (\ref{relation-gammas}) and (\ref{ave-series-perturbation}), we get
\eqa{ \label{ave-delta-F}
\expval{\delta F}_{\eta_\chi} &= \expval{\delta F}_\eta + \expval{\delta \dot{F} \,\delta\eta} + 3\Hu\,\expval{\delta F \delta\eta} \\[3pt]
&\hspace{30pt} - 3\Hu\,\expval{\delta F}\expval{\delta\eta} \,.
}
In particular, applying this general result to $\delta\eta$, we find
\eq{ \label{ave-delta-eta}
\expval{\delta \eta}_{\eta_\chi} = \expval{\delta \eta}_\eta + \expval{\delta \dot{\eta} \,\delta\eta} + 3\Hu\,\expval{(\delta\eta)^2} - 3\Hu\,\expval{\delta\eta}^2 \,.
}

Hence, the average value of perturbations on slightly different surfaces agree with each other at the first order, but generally differ at higher orders because of the small difference between the surfaces. We can now return to the main problem at hand, and focus on the function $F$ itself. When evaluated on the surface of constant fluid property $\chi$, its background value can be expanded in series as
\eqa{
F_0(\eta)\big|_{\eta_\chi = \eta_0} &= F_0(\eta_\chi + \delta\eta)\big|_{\eta_\chi = \eta_0} \\[3pt]
&= F_0 + \dot{F}_0\,\delta\eta|_{\eta_\chi = \eta_0} + \frac12 \ddot{F}_0\,(\delta\eta)^2\,,
}
where $F_0 = F_0(\eta_0)$, and similarly for the time derivatives $\dot{F}_0$ and $\ddot{F}_0$. Using (\ref{def-function-F}), the value of the function $F$ on that same surface is then
\eqa{
\frac{F}{F_0}\bigg|_{\eta_\chi = \eta_0} &= 1 + \delta F|_{\eta_\chi = \eta_0} + \frac{\dot{F}_0}{F_0}\,\delta\eta|_{\eta_\chi = \eta_0} \\[3pt]
&\hspace{20pt} + \frac{\dot{F}_0}{F_0}\,\delta F \delta\eta + \frac{\ddot{F}_0}{2F_0}\,(\delta\eta)^2\,.
}

Taking the average of this equation on the surface of constant fluid property $\chi$, and then using (\ref{ave-delta-F}) and (\ref{ave-delta-eta}), we finally get
\eqa{
\frac{\expval{F}_{\eta_\chi} - \expval{F}_\eta}{F_0} &= \frac{\dot{F}_0}{F_0} \,\expval{\delta\eta}_{\eta_\chi} + \frac{\ddot{F}_0}{2F_0}\,\expval{(\delta\eta)^2} + \expval{\delta \dot{F}\,\delta\eta} \\[3pt]
&\hspace{10pt} + \Big(\frac{\dot{F}_0}{F_0} + 3\Hu\Big)\expval{\delta F \delta\eta} - 3\Hu\,\expval{\delta F}\expval{\delta\eta} \,. 
}

Let's now apply this general result to the expansion rate of the spacetime, $F=H$. As usual, its background value $H_0$ is given by (\ref{background-expansion-rate}). Moreover, as shown in (\ref{linear-correction-vanishes}), the linear correction to the expansion rate vanishes when it is averaged over  surfaces of constant total density. Hence, with $\expval{\delta H} = 0$ at the first order, we get
\eqa{ \label{diff-two-averages}
\frac{\expval{H}_{\eta_\chi} - \expval{H}_\eta}{H_0} &= \frac{\dot{\Hu} - \Hu^2}{\Hu} \expval{\delta\eta}_{\eta_\chi} \,+\, \frac{2\Hu^2 + \dot{\Hu}}{\Hu}\expval{\delta H \,\delta\eta} \\[4pt]
&\hspace{10pt} +\, \expval{\delta \dot{H}\,\delta\eta} \,+\, \frac{\ddot{\Hu} - 3\Hu\dot{\Hu} + \Hu^3}{2\Hu}\,\expval{(\delta\eta)^2}\,.
}

This is the important result of this section. In particular, if we assume that the perturbation $\delta\eta$ vanishes on average over surfaces of constant total density, $\expval{\delta\eta}_\eta = 0$, its average (\ref{ave-delta-eta}) over the surfaces of constant $\chi$ becomes
\eq{
\expval{\delta \eta}_{\eta_\chi} = \expval{\delta \dot{\eta} \,\delta\eta} + 3\Hu\,\expval{(\delta\eta)^2}\,, 
}
which implies that
\eqa{
\frac{\expval{H}_{\eta_\chi} - \expval{H}_\eta}{H_0} &= \frac{2\Hu^2 + \dot{\Hu}}{\Hu}\expval{\delta H \,\delta\eta} + \expval{\delta \dot{H}\,\delta\eta} \\[3pt]
&\hspace{10pt} +  \frac{\ddot{\Hu} + 3\Hu\dot{\Hu} - 5\Hu^3}{2\Hu}\,\expval{(\delta\eta)^2} \\[3pt]
&\hspace{10pt} + \frac{\dot{\Hu} - \Hu^2}{\Hu}\, \expval{\delta\dot{\eta}\,\delta\eta}\,.
}

The long-wavelength limit of this result has been discussed recently in \cite{Comeau:2023a}, for the particular case of a spacetime containing two fluids with constant equations of state. In this paper, the density of the one of the fluids plays the role of the scalar property $\chi$.

\para{Partial density as cosmic clock}

So far in this section, the fluid property $\chi$ used as a cosmic clock has been kept completely general. Let's now consider the particular case when such a property is taken to be the density of some of the fluids contained in the spacetime. 

In that case, the perturbation $\delta\eta$ can be related to the entropy perturbation induced by the fact that there are two types of fluids in the spacetime, namely, the ones whose density is used as a cosmic clock, and the others which do not play that role. It is convenient to label the former as being of type $A$, and the latter as being of type $B$. 

Let $\rho_A$ and $\rho_B$ denote the total density and pressure of the fluids of type $A$, and $\rho_B$ and $p_B$ the corresponding quantities for the fluids of type $B$. The total density and pressure for all the fluids are then
\eqa{
&\rho = \rho_A + \rho_B\,, \\[4pt]
&p = p_A + p_B\,.
}

Each fluid category can comprise various components, with different properties and equations of state. Therefore, the pressure perturbations in each case can be written as the sum of an adiabatic contribution, proportional to the density perturbation, plus an entropy perturbation. At the first order, we have 
\eqa{
\delta p_A &= c_A^2 \,\delta \rho_A + \delta\mathcal{P}_A\,, \\[4pt]
\delta p_B &= c_B^2 \,\delta \rho_B + \delta\mathcal{P}_B\,,
}
where $c_A$ and $c_B$ denote the speed of sound in the fluids of types $A$ and $B$, respectively. Here, $\delta\mathcal{P}_A$ and $\delta\mathcal{P}_B$ represent the non-adiabatic perturbations which arise whenever the pressure does not depend exclusively on the density, but also on other variables, such as the individual densities of the fluid components. These additional variables can be labelled generically as representing ``entropy''.  

In the uniform-density gauge, the total density of all fluids is equal to its background value, $\delta\rho = 0$, which implies that $\delta\rho_B = -\delta\rho_A$. In that case, the total pressure perturbation is entirely non-adiabatic, and is given by
\eq{
\delta p = \delta\mathcal{P}_A + \delta\mathcal{P}_B + \delta\mathcal{P}_{AB}\,,
}
where $\delta\mathcal{P}_{AB}$ represents the non-adiabatic pressure perturbation generated by the presence of two types of fluids in the spacetime,
\eq{
\delta\mathcal{P}_{AB} = (c_A^2 - c_B^2)\,\delta\rho_A\,,
}
that is, using (\ref{eta-density}),
\eq{
\delta\eta = \frac{\delta\mathcal{P}_{AB}}{(c_B^2 - c_A^2)\,\dot{\rho}_A}\,.
}

Therefore, assuming the speeds of sound are different, the perturbation $\delta\eta$ can be expressed in terms of the entropy perturbation induced by the presence of two types of fluids in the spacetime. If the perturbations are purely adiabatic, then $\delta\eta = 0$, and vice-versa. This conclusion actually holds not only at the first order, as we just showed, but at all perturbative orders. 

Our original intention was to measure the average value of the expansion rate over a surface where the total density of the fluids of type $A$ is constant. This average value differs from the one over a surface where the total density of all fluids is constant by an amount related to $\delta\eta$. Hence, for the two averages to differ, $\delta\eta$ must not vanish, and so the fluid perturbations must not be purely adiabatic. In other words, the two averages coincide with each other when there are no entropy perturbations.

\section*{Acknowledgments}

The work done for this paper was supported in part by the Fonds de recherche du Qu\'ebec (FRQNT). I have also benefited greatly from discussions with my PhD advisor Robert Brandenberger (McGill University).

\bibliographystyle{mybibstyle}
\bibliography{references_back-reaction}

\providecommand{\href}[2]{#2}\begingroup\raggedright\begin{thebibliography}{10}

\bibitem{Rasanen:2003}
S.~Rasanen, ``{Dark energy from back-reaction},''  \textit{JCAP\/}
  \textbf{02\/}, p. 003, 2004,
  \href{https://arxiv.org/abs/astro-ph/0311257}{{\ttfamily astro-ph/0311257}}.

\bibitem{Barausse:2005}
E.~Barausse, S.~Matarrese and A.~Riotto, ``{The effect of inhomogeneities on
  the luminosity distance-redshift relation: Is dark energy necessary in a
  perturbed Universe?},''  \textit{Phys. Rev. D\/} \textbf{71\/}, p. 063537,
  2005, \href{https://arxiv.org/abs/astro-ph/0501152}{{\ttfamily
  astro-ph/0501152}}.

\bibitem{Geshnizjani:2005}
G.~Geshnizjani, D.J.H.~Chung and N.~Afshordi, ``{Do large-scale inhomogeneities
  explain away dark energy?},''  \textit{Phys. Rev. D\/} \textbf{72\/}, p.
  023517, 2005, \href{https://arxiv.org/abs/astro-ph/0503553}{{\ttfamily
  astro-ph/0503553}}.

\bibitem{Hirata:2005}
C.M.~Hirata and U.~Seljak, ``{Can super-Horizon cosmological perturbations
  explain the acceleration of the Universe?},''  \textit{Phys. Rev. D\/}
  \textbf{72\/}, p. 083501, 2005,
  \href{https://arxiv.org/abs/astro-ph/0503582}{{\ttfamily astro-ph/0503582}}.

\bibitem{Brandenberger:2002}
R.~Brandenberger, ``{Back-reaction of cosmological perturbations and the
  cosmological constant problem},''  2002,
  \href{https://arxiv.org/abs/hep-th/0210165}{{\ttfamily hep-th/0210165}}.

\bibitem{Tsamis:1996a}
N.C.~Tsamis and R.P.~Woodard, ``Quantum gravity slows inflation,''
  \textit{Nucl. Phys. B\/} \textbf{474\/}, p. 235, 1996,
  \href{https://arxiv.org/abs/hep-ph/9602315}{{\ttfamily hep-ph/9602315}}.

\bibitem{Tsamis:1996b}
N.C.~Tsamis and R.P.~Woodard, ``The quantum gravitational back reaction on
  inflation,''  \textit{Annals Phys.\/} \textbf{253\/}, p. 1, 1997,
  \href{https://arxiv.org/abs/hep-ph/9602316}{{\ttfamily hep-ph/9602316}}.

\bibitem{Kolb:2004}
E.W.~Kolb, S.~Matarrese, A.~Notari and A.~Riotto, ``{The effect of
  inhomogeneities on the expansion rate of the universe},''  \textit{Phys. Rev.
  D\/} \textbf{71\/}, p. 023524, 2005,
  \href{https://arxiv.org/abs/hep-ph/0409038}{{\ttfamily hep-ph/0409038}}.

\bibitem{Geshnizjani:2002}
G.~Geshnizjani and R.~Brandenberger, ``Back-reaction and local cosmological
  expansion rate,''  \textit{Phys. Rev. D\/} \textbf{66\/}, p. 123507, 2002,
  \href{https://arxiv.org/abs/gr-qc/0204074}{{\ttfamily gr-qc/0204074}}.

\bibitem{Geshnizjani:2003}
G.~Geshnizjani and R.~Brandenberger, ``{Back-reaction of perturbations in two
  scalar field inflationary models},''  \textit{JCAP\/} \textbf{04\/}, p. 006,
  2005, \href{https://arxiv.org/abs/hep-th/0310265}{{\ttfamily
  hep-th/0310265}}.

\bibitem{Buchert:1999}
T.~Buchert, ``{On average properties of inhomogeneous fluids in general
  relativity: Dust cosmologies},''  \textit{Gen. Rel. Grav.\/} \textbf{32\/},
  p. 105, 2000, \href{https://arxiv.org/abs/gr-qc/9906015}{{\ttfamily
  gr-qc/9906015}}.

\bibitem{Buchert:2001}
T.~Buchert, ``{On average properties of inhomogeneous fluids in general
  relativity: Perfect fluid cosmologies},''  \textit{Gen. Rel. Grav.\/}
  \textbf{33\/}, p. 1381, 2001,
  \href{https://arxiv.org/abs/gr-qc/0102049}{{\ttfamily gr-qc/0102049}}.

\bibitem{Abramo:1997}
L.~Abramo, R.~Brandenberger and V.F.~Mukhanov, ``{The energy-momentum tensor
  for cosmological perturbations},''  \textit{Phys. Rev. D\/} \textbf{56\/}, p.
  3248, 1997, \href{https://arxiv.org/abs/gr-qc/9704037}{{\ttfamily
  gr-qc/9704037}}.

\bibitem{Unruh:1998}
W.~Unruh, ``Cosmological long-wavelength perturbations,''  1998,
  \href{https://arxiv.org/abs/astro-ph/9802323}{{\ttfamily astro-ph/9802323}}.

\bibitem{Abramo:1998}
L.~Abramo and R.P.~Woodard, ``One loop back-reaction on chaotic inflation,''
  \textit{Phys. Rev. D\/} \textbf{60\/}, p. 044010, 1999,
  \href{https://arxiv.org/abs/astro-ph/9811430}{{\ttfamily astro-ph/9811430}}.

\bibitem{Afshordi:2000}
N.~Afshordi and R.~Brandenberger, ``Super Hubble nonlinear perturbations during
  inflation,''  \textit{Phys. Rev. D\/} \textbf{63\/}, p. 123505, 2001,
  \href{https://arxiv.org/abs/gr-qc/0011075}{{\ttfamily gr-qc/0011075}}.

\bibitem{Adamek:2017}
J.~Adamek, C.~Clarkson, D.~Daverio, R.~Durrer and M.~Kunz, ``{Safely smoothing
  spacetime: back-reaction in relativistic cosmological simulations},''
  \textit{Class. Quant. Grav.\/} \textbf{36\/}, p. 014001, 2019,
  \href{https://arxiv.org/abs/1706.09309}{{\ttfamily 1706.09309}}.

\bibitem{Comeau:2023a}
V.~Comeau and R.~Brandenberger, ``{Back-reaction of long-wavelength
  cosmological fluctuations as measured by a clock field},''  2023,
  \href{https://arxiv.org/abs/2302.05873}{{\ttfamily 2302.05873}}.

\bibitem{Marozzi:2012}
G.~Marozzi, G.P.~Vacca and R.~Brandenberger, ``Cosmological back-reaction for a
  test field observer in a chaotic inflationary model,''  \textit{JCAP\/}
  \textbf{02\/}, p. 027, 2013,
  \href{https://arxiv.org/abs/1212.6029}{{\ttfamily 1212.6029}}.

\bibitem{Brandenberger:2018}
R.~Brandenberger, L.L.~Graef, G.~Marozzi and G.P.~Vacca, ``Back-reaction of
  super-Hubble cosmological perturbations beyond perturbation theory,''
  \textit{Phys. Rev. D\/} \textbf{98\/}, p. 103523, 2018,
  \href{https://arxiv.org/abs/1807.07494}{{\ttfamily 1807.07494}}.

\bibitem{Ehlers:1993}
J.~Ehlers, ``Contributions to the relativistic mechanics of continuous media,''
   \textit{Gen. Rel. Grav.\/} \textbf{25\/}, p. 1225, 1993.

\bibitem{Mukhanov:1992}
V.F.~Mukhanov, H.A.~Feldman and R.~Brandenberger, ``Theory of cosmological
  perturbations,''  \textit{Physics Reports\/} \textbf{215\/}, p. 203, 1992.

\bibitem{Brandenberger:2003}
R.~Brandenberger, ``Lectures on the theory of cosmological perturbations,''
  \textit{Lect. Notes Phys.\/} \textbf{646\/}, p. 127, 2004,
  \href{https://arxiv.org/abs/hep-th/0306071}{{\ttfamily hep-th/0306071}}.

\bibitem{Andersson:2006}
N.~Andersson and G.L.~Comer, ``{Relativistic fluid dynamics: Physics for many
  different scales},''  \textit{Living Rev. Rel.\/} \textbf{10\/}, p. 1, 2007,
  \href{https://arxiv.org/abs/gr-qc/0605010}{{\ttfamily gr-qc/0605010}}.

\bibitem{Rezzolla:2013}
L.~Rezzolla and O.~Zanotti, \emph{{Relativistic Hydrodynamics}}, Oxford
  University Press, 2013.

\bibitem{Blachier:2023}
B.~Blachier, P.~Auclair, C.~Ringeval and V.~Vennin, ``{Spatial curvature from
  super-Hubble cosmological fluctuations},''  2023,
  \href{https://arxiv.org/abs/2302.14530}{{\ttfamily 2302.14530}}.

\end{thebibliography}\endgroup

\end{document}